\pdfoutput=1

\documentclass[aps,pra,preprint,superscriptaddress, longbibliography]{revtex4-1}
\usepackage{graphicx}
\usepackage[usenames]{xcolor}
\usepackage{bm}

\begin{document}

% Use the \preprint command to place your local institutional report
% number in the upper righthand corner of the title page in preprint mode.
% Multiple \preprint commands are allowed.
% Use the 'preprintnumbers' class option to override journal defaults
% to display numbers if necessary
%\preprint{}

%Title of paper
\title{Small slot waveguide rings for on-chip quantum optical circuits}

% repeat the \author .. \affiliation  etc. as needed
% \email, \thanks, \homepage, \altaffiliation all apply to the current
% author. Explanatory text should go in the []'s, actual e-mail
% address or url should go in the {}'s for \email and \homepage.
% Please use the appropriate macro foreach each type of information

% \affiliation command applies to all authors since the last
% \affiliation command. The \affiliation command should follow the
% other information
% \affiliation can be followed by \email, \homepage, \thanks as well.
\author{Nir Rotenberg}
\author{Pierre T\"{u}rschmann}
\author{Harald R. Haakh}
\author{Diego Martin-Cano}
%\email[]{Your e-mail address}
%\homepage[]{Your web page}
%\thanks{}
%\altaffiliation{}
\affiliation{Max Planck Institute for the Science of Light (MPL), D-91058 Erlangen, Germany}
\author{Stephan G\"{o}tzinger}
\affiliation{Department of Physics, Friedrich Alexander University of Erlangen-N{\"u}rnberg (FAU), D-91058 Erlangen, Germany}
\affiliation{Max Planck Institute for the Science of Light (MPL), D-91058 Erlangen, Germany}
\affiliation{2Graduate School in Advanced Optical Technologies (SAOT), Friedrich Alexander University (FAU) Erlangen-N{\"u}rnberg, 91052 Erlangen, Germany}
\author{Vahid Sandoghdar}
\affiliation{Max Planck Institute for the Science of Light (MPL), D-91058 Erlangen, Germany}
\affiliation{Department of Physics, Friedrich Alexander University of Erlangen-N{\"u}rnberg (FAU), D-91058 Erlangen, Germany}

%Collaboration name if desired (requires use of superscriptaddress
%option in \documentclass). \noaffiliation is required (may also be
%used with the \author command).
%\collaboration can be followed by \email, \homepage, \thanks as well.
%\collaboration{}
%\noaffiliation

\date{\today}

\begin{abstract}
Nanophotonic interfaces between single emitters and light promise to enable new quantum optical technologies. Here, we use a combination of finite element simulations and analytic quantum theory to investigate the interaction of various quantum emitters with slot-waveguide rings. We predict that for rings with radii as small as 1.44~$\mu$m (Q = 27,900), near-unity emitter-waveguide coupling efficiencies and emission enhancements on the order of 1300 can be achieved.  By tuning the ring geometry or introducing losses, we show that realistic emitter-ring systems can be made to be either weakly or strongly coupled, so that we can observe Rabi oscillations in the decay dynamics even for micron-sized rings.  Moreover, we demonstrate that slot waveguide rings can be used to directionally couple emission, again with near-unity efficiency.  Our results pave the way for integrated solid-state quantum circuits involving various emitters.
\end{abstract}

% insert suggested PACS numbers in braces on next line
% insert suggested keywords - APS authors don't need to do this
%\keywords{}

%\maketitle must follow title, authors, abstract, \pacs, and \keywords
\maketitle

% body of paper here - Use proper section commands
% References should be done using the \cite, \ref, and \label commands
\section{Introduction}
The cutting edge of solid-state quantum optics is determined by our ability to realize phenomena such as photon-mediated cooperative effects between multiple quantum emitters~\cite{QuantumPhases, PhotonGasI, PhotonGasII, LuttingerLiquid}, few-photon nonlinearities~\cite{FewPhoton_NLO_QD, FewPhoton_NLO_Molecule, Squeezing}, and strong light-matter coupling~\cite{Strong_Coupling_I, Strong_Coupling_II}, which provide the resources that are crucial for scalable quantum simulation and information processing~\cite{QuantumSimulator, QDsPhCWs, QuantumGates}. An important bottleneck in this endeavor is the efficient coupling of light and matter.  Ideally, one would like a single photon to interact with a single quantum emitter such as a quantum dot, color center, ion, or molecule with 100\% efficiency.

Traditionally, cavities have been used to boost light-matter coupling by significantly increasing the time during which a photon and an emitter interact. A variety of resonators such as bulk Fabry-P{\'e}rot cavities, microspheres, microdisks, micropillars, or photonic crystal cavities can retain a photon for millions of optical cycles, increasing the probability $\left(\beta\right)$ that a photon interacts with a quantum emitter to unity~\cite{Microcavities_Review}. In other words, a photon is always emitted into the cavity mode. The high quality factors (Q) required for this operation, however, usually pose severe technical challenges. In particular, it becomes imperative that one tunes the narrow resonances of the cavity and the emitter to each other and stabilizes the cavity length to down to a few picometers \cite{Cavity_Stabilization}.  Simultaneous coupling of several emitters or frequencies that can be individually addressed is, therefore, not within reach.

As an alternative to cavity-enhanced interaction, several groups have investigated single-pass coupling via near-field optics \cite{Gerhardt:07}, tight focusing \cite{Vamivakas:07, Wrigge:08, Tey:08, Streed:12} or a subwavelength waveguide (nanoguide) \cite{Shen:05, Vetsch:10, Yalla:12, Molecules_Nanoguide}. The key concept in this approach is spatial mode matching between the photon and the emitter radiation pattern. Although the coupling efficiency $\beta$ can theoretically reach unity \cite{Zumofen:08}, it is a challenge to identify well-behaved coherent transitions known in atomic physics in the solid state. Here, issues such as the quantum efficiency, phonon dephasing, or lossy transitions limit the scattering cross section of a given transition. For example, in the case of organic dye molecules the Frank-Condon and Debye-Waller factors reduce the overall efficiency by about $50-70\%$ \cite{Gerhardt:07} while for nitrogen-vacancy centers in diamond, strong phonon wings and the quantum efficiency limit the efficiency to well below $10\%$~\cite{QO_NV_Review}. In cavity-coupling, one can hope to compensate for such photophysical deficiencies by strong enhancement of the interaction between the cavity mode and the emitter~\cite{Nanophotonics}.

The central advantage of a cavity-free coupling is its immense bandwidth. In this work, we provide an example, where the advantages of single-pass and cavity couplings are combined through the design of feedback geometries with moderate $Q$. Of the different cavity-free approaches, the nanoguide geometry is particularly attractive for this purpose because it can be implemented on a chip and be used as the building block of quantum optical circuits. Moderate cavity feedback would also be particularly advantages for this platform because it is otherwise a great challenge to achieve a very large index contrasts between the nanoguide and its surrounding, necessary for reaching high $\beta$ factors. So far, this issue has been addressed via slow light photonic crystals ~\cite{QO_Plasmonic_Waveguides, NearUnityCoupling_PhCWs}. Another proposal has been to use slot waveguides \cite{QO_SWsI, QO_SWsII}.

\begin{figure*}[!tp]
  \centering
  \includegraphics[width=13cm]{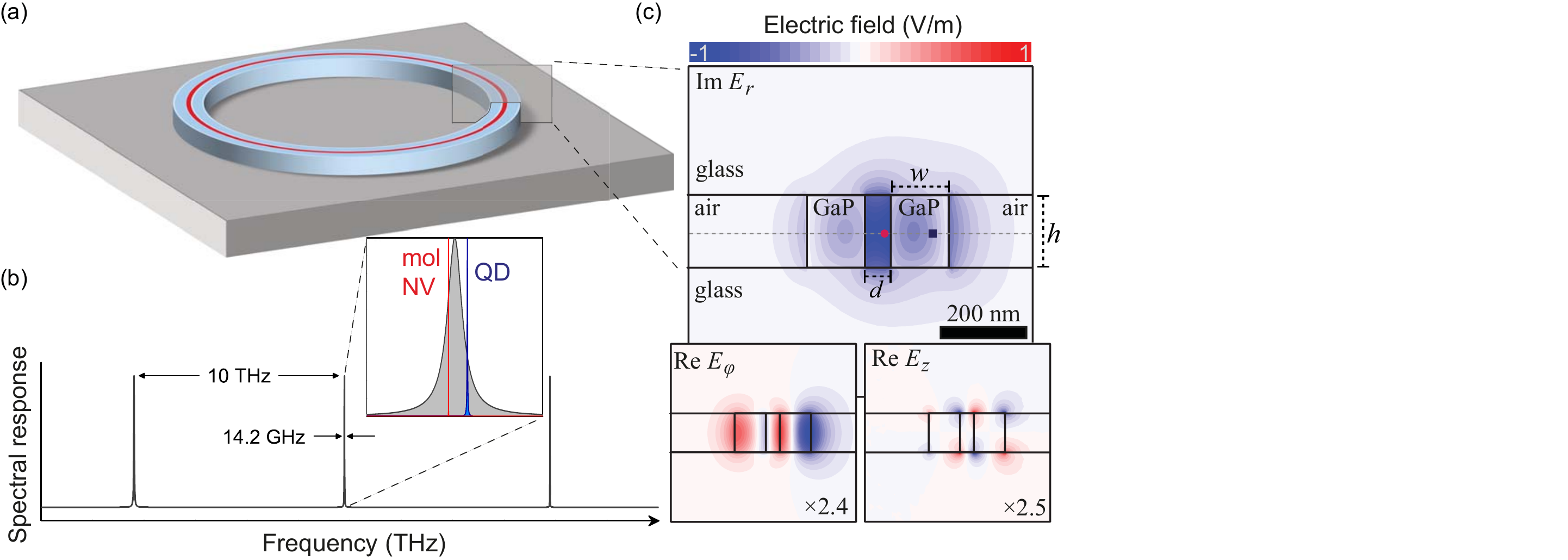}
  \caption{Slot waveguide rings for quantum optics. (a) Sketch of the slot waveguide ring, which is comprised of two high index dielectric bars (here, GaP with $n = 3.2$) that are separated by a low index material.  This ring is sandwiched between two glass plates of $n = 1.48$ (here, we only show the bottom substrate, for clarity), and has air pockets to each side.  (b) The spectral response of a slot waveguide ring with height $h = 175$~nm, width $w = 135$~nm, gap size $d = 60$~nm and radius $r = 1.44$~$\mu$m, as shown in (c).  The central mode, corresponding to the $24^{th}$ order TE-polarized azimuthal mode, whose field profile is shown in (c), has a bandwidth of 14.2 GHz, and a free spectral range of 10 THz.  The zoom in, shows this mode relative to the zero phonon line resonances of single molecules and NV centers in diamond (red curve, $\approx 10 - 30$~MHz bandwidth) and epitaxially grown quantum dots (blue curve, 400 MHz bandwidth) at cryogenic temperatures.  In (c), the dominant, radial component is shown in the main figure pane, while the azimuthal and out-of-plane components are shown in the sub-panels.  The relative scaling of the different components are shown in the bottom right corners (e.g. the maximum of $E_r$ is 2.4 times that of $E_{\varphi}$).  The (red) circle shows a favorable position for a quantum emitter, such as an organic molecule, which is embedded a low index dielectric, while the (blue) square is a favorable position for an emitter such as a quantum dot or NV center which is embedded in a high index material.}\label{fig:System}
\end{figure*}

Here, we begin with a slot waveguide that couples well to single quantum emitters $\left(\beta \approx 0.6\right)$ and bend it into a ring, as shown in Fig.~\ref{fig:System}a. We model a realistic implementation of the slot-waveguide ring compatible with nanofabrication capabilities. By choosing a small radius $r < 1.5 \, \mu$m, we minimize the structure footprint, while keeping the balance between the $Q$ $\left(>3000\right)$ of the ring and its performance as a quantum optical platform.  As we show, this system maintains the broadband nature of waveguides relative to lifetime-limited transitions of solid-state emitter (Fig.~\ref{fig:System}b) and allows for near-unity $\beta$ and even to enter into the strong coupling regime.  We conclude by considering two specific implementations of our system: one where the emitter is sitting inside the slot, as would be the case for single organic molecules or colloidal quantum dots, and the other for an emitter such as a quantum dot or an NV center embedded in one of the high index bars of the slot waveguide (see Fig.~\ref{fig:System}c). In the latter, we explore the possibility of chiral-emission where the state of the emitter determines in which direction a photon is emitted.  In either case, the robust performance of the slot-waveguide ring suggest that it is a powerful platform for future quantum nanophotonic experiments and applications.

\section{Quantum optics in slot-waveguide rings}\label{sec:SWg_QO}
\subsection{Solid-state quantum emitters in linear nanoguides}
Nanoscale waveguides (nanoguides) provide a flexible and scalable platform for quantum optics. The simplest version of a nanoguide consists of a nanoscopic rectangular channel that is surrounded by a lower refractive-index medium, and the efficiency with which it couples to emission depends on the magnitude of this refractive index mismatch.  Simply put, a greater refractive index difference between the core and the surrounding results in larger field enhancement inside the nanoguide, and a more efficient interface with a quantum emitter. The optical properties of nanoguides depend on their geometry, meaning that mode profiles, light confinement and bandwidths are all easily tuned. This versatility ensures that nanoguides can be designed to interface with the many different quantum emitters, each of which has different optical properties and is suited for different applications~\cite{Single_photon_sources}. In practice, however, the coupling efficiency of the system is limited, as its constituent materials are often predetermined by the type of emitter used.

Quantum emitters, in general, act as two (or three) level systems. A transition between these levels occurs as a result of a charge redistribution that can be described by a transition dipole $\mathbf{d}_{e}$, and is accompanied by the absorption or emission of a photon of angular frequency $\omega_e$.  Each transition resonance is described by a homogenous linewidth $\gamma_{\rm{hom}}$ (see Fig.~\ref{fig:System}b, for a comparison of emitter linewidth relative to a cavity resonance) that is typically distributed over an inhomogeneous spectrum that is typically greater than 1 THz in the solid state. This inhomogeneous broadening arises because each emitter experiences a slightly different local environment within its dielectric host matrix.

Integration of various solid-state emitters in waveguides requires considerations specific to each system. In particular, some emitters such as single organic molecules or colloidal quantum dots are generally embedded in a fairly low index dielectric. Thus, a nanoguide made of the host matrix results in low coupling efficiencies around 0.1 to 0.2~\cite{Molecules_Nanoguide}, while inclusion of single molecules or colloidal quantum dots into standard nanophotonic structures fabricated out of high-$n$ dielectrics is not trivial either. Rare earth ions, epitaxially grown quantum dots, or vacancy centers in diamond, on the other hand, are inherently embedded in high-$n$ matrices that can form nanoguides; The coupling efficiency typically remains below 0.75, though it can be increased by introducing resonances to the structure~\cite{NV_Rings, Diamond_Rings}.

\subsection{Straight slot waveguides}
Here, we turn to a slot waveguide~\cite{QO_SWsI, QO_SWsII, QO_SWsIII}, whose cross-cut is outlined in Fig.~\ref{fig:System}c, showing an ultrathin low $n$ region sandwiched between two higher index channels.  This geometry leads to an effective coupling of the modes of the two high index channels, resulting in confinement of the propagating mode to the low index region, which is highly subwavelength in size.  As a best case, we calculate that an emitter in a 60 nm wide, $n = 1.6$ dielectric between two gallium phosphide (GaP) channels $\left(n = 3.2\right)$, will experience a coupling efficiency $\beta = 0.75$ and an enhancement of the emission $\chi = 3.25$ (see Appendix~\ref{sec:etaBeta} for the calculation of $\beta$ and $\chi$).  These correspond to more than 4-fold increase of $\beta$ and 3-fold increase of $\chi$ relative to a simple nanoguide geometry~\cite{Molecules_Nanoguide}.  The broadband nature of the waveguides results in enhancement of the resonant florescence at $\omega_e$, but equally of all red-shifted emission.

\subsection{Emission into slot waveguide rings}\label{sec:EmissionSWR}
By bending the slot waveguide into a ring, as shown schematically in Fig.~\ref{fig:System}a, we can add optical feedback to our system and increase $\beta$ from 0.75 towards unity.  The spectral response of a 1.44~$\mu$m radius ring, where bending losses dominate (see Sec.~\ref{sec:ImpNMat}), is shown in Fig.~\ref{fig:System}b (other dimensions given in the caption); here, we observe narrow modes separated by 10 THz.  A zoom onto the 24$^{th}$ order mode reveals a calculated spectral full-width at half maximum (FWHM) of 14.2 GHz, which is up to 1000 times broader than the natural linewidths of various typical emitter resonances.  As examples, in the inset to Fig.~\ref{fig:System}b we overlay zero-phonon line linewidths of single molecules~\cite{DBTinNaph} and vacancy centers in diamond~\cite{Linewidth_NV} (10-30 MHz FHWM, red curve) and of epitaxially grown quantum dots~\cite{Linewidth_QD} (530 MHz, blue curve) over the ring resonance.  For this ring geometry, the non-zero width of the resonances is due solely to bending losses, which result in complex eigenfrequencies. Here, for instance, the complex mode frequency is $\omega_{\rm cav} + i\gamma_{\rm cav} / 2= 2 \pi \left(3.947\times10^{14} + i7.082\times 10^9\right)$~rad/s (calculated using the commercial eigenfrequency solver COMSOL).  These bending or radiative losses are the limiting factor of the quality of the resonator given by $Q = Q_{\rm rad} = \omega_{\rm cav} / \gamma_{\rm cav} = 27,900$.

Creating a resonator out of the slot waveguide also affects the optical eigenmode of the structure, albeit in a more subtle manner than the modification to its spectral response.  In contrast to a straight slot waveguide, the mode of the ring is slightly asymmetric, as we see in Fig.~\ref{fig:System}c.  This asymmetry is visible in all three field components, as the fields on the outside of the ring (right side) are slightly larger.  As is the case for a straight slot waveguide, the field is largest in the slot, where it is also primarily radially polarized.  This slot, then, is ideally suited for emitters with linear transition dipoles embedded in a low-$n$ dielectric.  A good position for such an emitter is marked by the red circle in Fig.~\ref{fig:System}c; note that this position is clearly off-center, due to the aforementioned mode asymmetry.  This slot waveguide ring is also a good platform for emitters that require a high-$n$ host dielectric.  These, be they epitaxially grown quantum dots or defect centers in diamond, could be placed at the position of the blue square in Fig.~\ref{fig:System}c (see Sec.~\ref{sec:Imp_highN} for a detailed explanation of why such a placement is advantageous).

\begin{figure}[!tp]
  \centering
  \includegraphics[width=8cm]{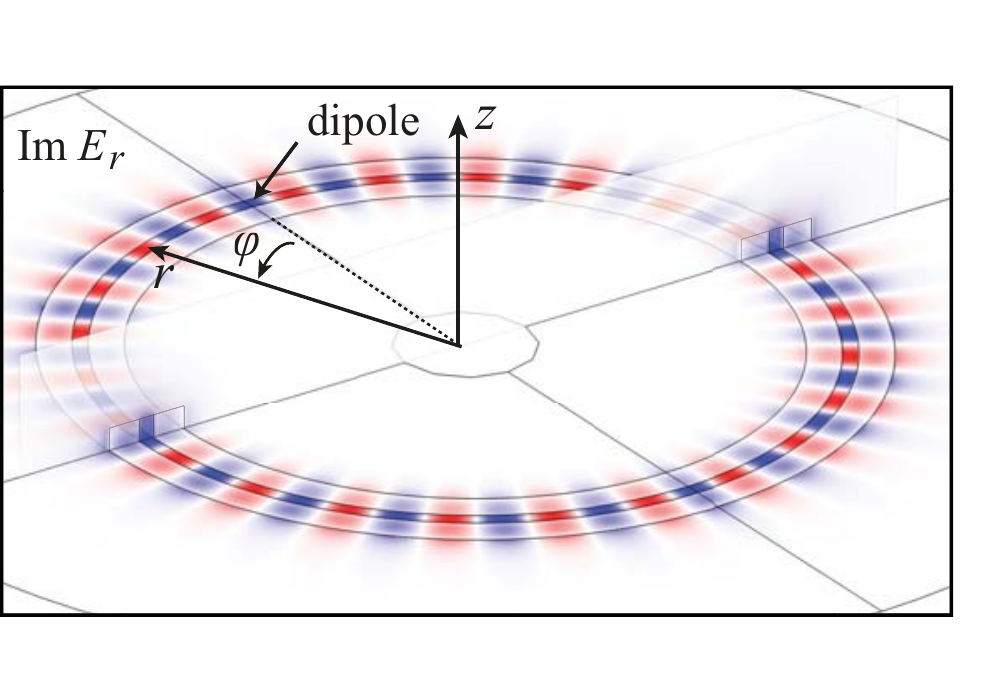}\\
  \caption{Emission into a slot waveguide ring.  A full three-dimensional calculation of the imaginary component of the radial electric field radiated by an emitter located in the slot (circle in Fig.~\ref{fig:System}c). The ring has a 1.44 $\mu$m radius, and the cylindrical coordinate axes of our system are shown.}\label{fig:Emission}
\end{figure}

To study the interaction of an emitter with the slot waveguide mode, we perform fully three-dimensional finite element method simulations of the ring structure. A radially oriented transition dipole that oscillates at the ring resonance frequency (i.e. $\omega_e = \omega_{\rm cav}$) represents the emitter, and is placed in the slot.  We then look at the steady-state field generated by this dipole, as shown in the $z = 0$ and $\varphi = \pi/2$ planes in Fig.~\ref{fig:Emission} (where $\varphi = 0$ at the position of the dipole), for a $1.44$~$\mu$m radius ring.  In this image the radiation is nearly fully coupled into the photonic modes of the ring. To determine $\chi$ quantitatively, we use the field map, as explained in Appendix~\ref{sec:etaBeta} and compute the fraction of the power radiated by a dipole into our nanophotonic system.  For the $1.44$~$\mu$m ring, we calculate that $\chi = 1,330$, a 400-fold enhancement of emission with respect to an identical emitter in a straight slot waveguide.  Similarly, we calculate $\beta$ by comparing the total emitted power to the one found far along the waveguide.  We find $\beta=0.995$, meaning that only 1/200 photons leaks out of the waveguide as compared to 1/4 in a straight nanoguide.

A similar analysis reveals remarkably efficient couplings between the ring and quantum emitters even when they are placed well away from the mode maximum in the slot.  An emitter embedded in the high-$n$ channel, as shown by the square symbol in Fig.~\ref{fig:System}c, would experience $\chi = 56$ and $\beta = 0.99$ (see Sec.~\ref{sec:Imp_highN} for more details).  That is, slot waveguide rings are compatible with all manner of quantum emitters, including quantum dots or NV centers, which would necessarily be placed out of the field maximum.

In addition to the improvement of the spatial coupling to a single mode, the slot waveguide ring will favor the resonant emission on the zero-phonon line compared to the Stokes-shifted coupling to vibrational states or phonon wings, thus, improving the branching ratio (oscillator strength) of solid-state emitters. This renders a solid-state emitter inside such a ring  like an ideal two-level emitter such as an atom with an overall outcome of $\beta\simeq1$. Clearly, even a ring with a geometric cross-section of only 6.5~$\mu$m$^2$ and a moderate $Q = 27900$ acts as a near-ideal interface between an emitter and photons.

\subsection{Emission dynamics}
We now turn to the dynamics of the emitter-waveguide resonator interaction.  In the framework of macroscopic quantum electrodynamics~\cite{Theory_dA2Gamma, Emitter_Cavity_Theory}, an emitter that is prepared in its excited state will decay according to (see Appendix~\ref{sec:Theory}),
\begin{equation}\label{eq:Ce}
    C_{e}\left(t\right)=\frac{e^{-\Gamma t/2}}{2D}\left[\left(D+\Gamma\right)e^{Dt/2}+\left(D-\Gamma\right)e^{-Dt/2}\right],
\end{equation}
where \begin{equation}\label{eq:D}
    D=\sqrt{\Gamma^{2} - K_{0}},
\end{equation}
and $K_{0}= \left(\omega_{\rm cav}^{2} /  \omega_{e}^{2}\right)  \chi \gamma_{\rm cav}\gamma_{\rm{hom}}$.  In these equations $\Gamma=i\left(\omega_{\rm cav}-\omega_{e}\right)+\gamma_{\rm cav}/2$ contains both the loss rate of our nanophotonic resonance and a phase due to detuning between the ring and emitter resonances.  In what follows, we assume $\omega_{\rm cav} = \omega_{e}$ and hence $\Gamma$ is simply the loss rate of our system.  It follows that $D$ depends on the difference between the rate at which energy is lost and the rate at which energy is exchanged between the emitter and the photonic mode.

\begin{figure}[!tp]
  \centering
  \includegraphics[width=8cm]{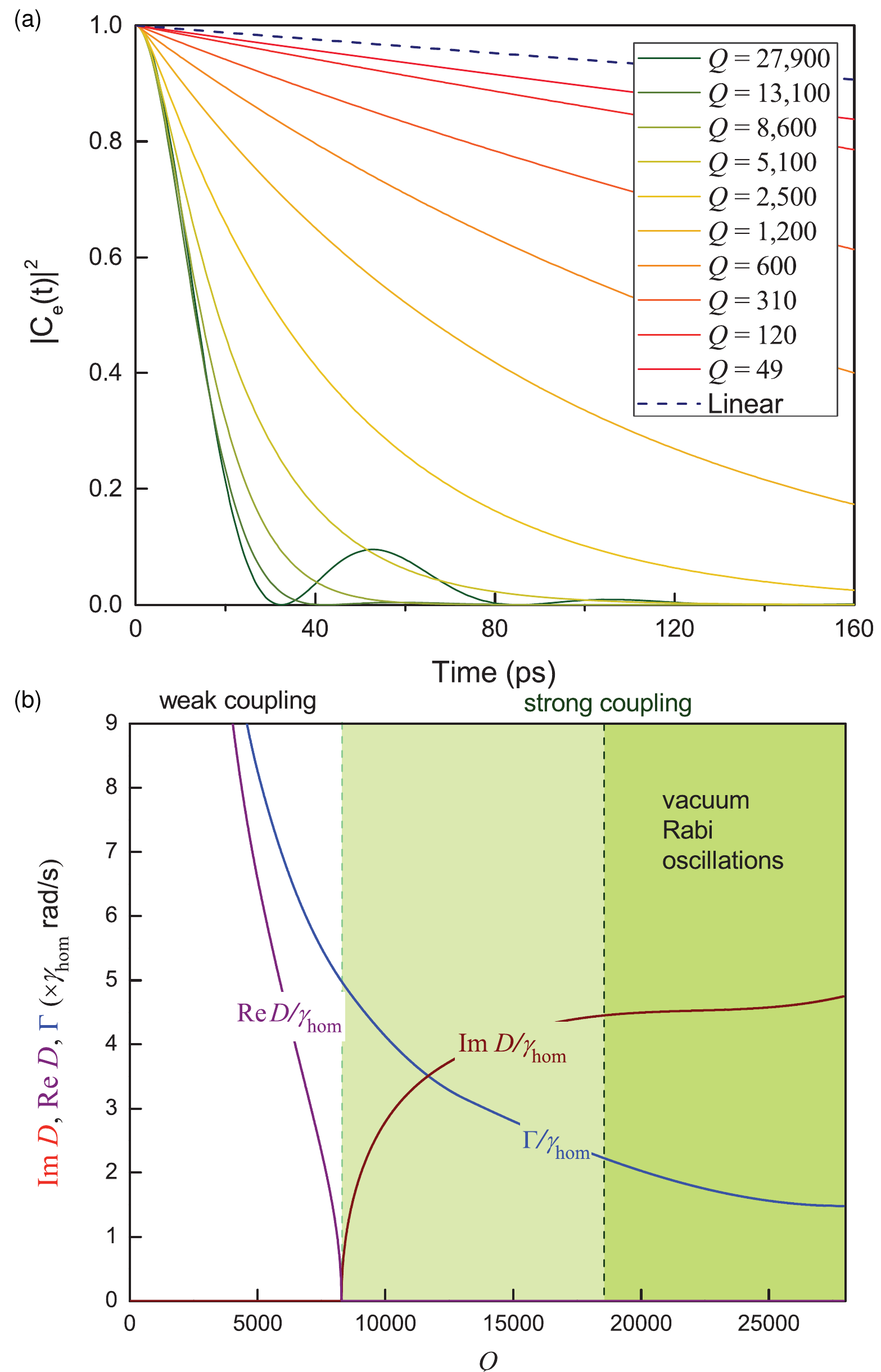}

  \caption{Decay dynamics of emitters coupled to ring resonators. (a) The time-dependent probability to find the emitter in its excited state for $Q$ ranging from 49 to 27900 for a 1.44~$\mu$m radius ring.  Rabi oscillations are clearly visible in the high $Q$ decay, while for small values of $Q$ the decay dynamics approach that of a lossless, straight slot waveguide (dashed curve).  (b) The different rates of the decay dynamics as a function of the resonance $Q$, normalized to $\gamma_{\rm{hom}}$.  The transition from the weak to strong coupling regimes occurs when the value of $D$ becomes imaginary.  In the strongly coupled regime, the region where Rabi oscillations can be observed (Im$D \geq 2\gamma$) is labeled, as is the low $Q$ region where a monoexponential decay of the emitter is observed.}\label{fig:Dynamics}
\end{figure}

The dynamics of the emitter's decay are determined by the interplay between $\Gamma$ and $D$, which, in turn, are dependent on the resonator quality factor, $Q$.  Equations~(\ref{eq:Ce}) and~(\ref{eq:D}) allow us to quantify these dependencies and to calculate the probability to find the emitter in its excited state $\left| C_{e}\left(t\right)\right|^2$.  The results of these calculations for rings with $Q$ ranging from 49 to 27900 are shown in Fig.~\ref{fig:Dynamics}(a) (see Sec.~\ref{sec:NanophotonicTuning} for an explanation on how $Q$ may be varied by introducing losses into the system).  Here, we take $\gamma_{\rm{hom}} = 30$~MHz, which is typical for a single organic molecule.

We observe markedly different dynamics for the different rings, ranging from a clear oscillatory behavior for the large $Q$ rings to a slow exponential decay for small $Q$ resonators.  In other words, the decay dynamics show that these slot waveguide rings can couple either strongly or weakly to quantum emitters, and the coupling strength can be tuned by varying the resonator $Q$.

We can, in fact, understand the different coupling regimes by considering Eqs.~(\ref{eq:Ce}) and~(\ref{eq:D}) and how $\Gamma$ and $D$ depend on $Q$ [Fig.~\ref{fig:Dynamics}b].  The transition from the strong to the weak coupling regimes occurs when $\Gamma^2 = K_0$, where $D$ evolves from being imaginary to real valued and $C_e\left(t\right)$ loses its oscillatory nature.  In our system, this change occurs for the moderate value of $Q = 8,300$.  For larger $Q$ resonators, when $\mbox{Im}D \geq 2\Gamma$, Rabi oscillations are clearly visible, as is the case in Fig.~\ref{fig:Dynamics}a.

In contrast, when $D$ is real valued, the emitter and resonator photons are weakly coupled and we observe exponential decay dynamics in Fig.~\ref{fig:Dynamics}a.  As expected, the decay constant approaches that of an emitter in a straight slot waveguide (dashed curve in Fig.~\ref{fig:Dynamics}a) as $Q$ decreases.  As we shall see below (Sec.~\ref{sec:Imp_LowN}), well into the regime where the resonator and emitter are weakly coupled (where $\chi > 200$, see Fig.~\ref{fig:BetaEta}), we still find $\beta\simeq1$, demonstrating the power of this quantum optical platform.

\section{Design considerations}
The previous sections laid out the response of an idealized slot-waveguide ring, namely one with no losses beyond those associated with the bending of the waveguide.  It is fair to ponder how such a ring would fare under more realistic conditions.  In this section, we answer this question, first examining the types of imperfections that can be reasonably expected for this type of nanophotonic structure and their consequences to the performance of the ring.

\subsection{Nanofabrication: Imperfections and materials}\label{sec:ImpNMat}
Nanofabrication techniques typically result in imperfections, leading to additional loss channels such as scattering losses due to surface roughness. In addition, depending on the choice of material, dopings or imperfections in the thin dielectric layers can cause absorption losses.  The total quality factor of the ring resonator can then be expressed as a sum of three contributions, $Q^{-1} = Q_{\rm rad}^{-1} + Q_{\rm scat}^{-1} + Q_{\rm abs}^{-1}$, corresponding to the radiative, scattering, and absorption channels. For a realistic performance, it is important to examine the competition among these contributions. We choose to neglect absorption losses in this work since several non-absorbing dielectric and semiconductor layers are available. We note, however, that it will be straightforward to extend our results to absorbing material because both absorption and scattering losses effect the optical properties of the ring by limiting the propagation length of the light.

The deposition and nanopatterning of thin films typically results in surface roughness that is on the order of 2-5 nm for granular films (such as TiO$_2$) and can be much smaller for single-crystalline or amorphous materials (such as GaP).  In the Rayleigh scattering model, $Q_{\rm scat}$ is inversely proportional to both the square of the root-mean-square (RMS) size of surface features and their correlation length~\cite{Q_SurfaceRoughness}.  A realistic value for the surface roughness of 2 nm, with a corresponding correlation length of 10~nm would  result in $Q_{\rm scat} = 2.1\times10^6$ at a wavelength of 760~nm.  In fact, it is only when the RMS roughness approaches 10 nm (and the correlation length 100 nm) that $Q_{\rm scat}$ drops below 20000.

We also note that although we have used GaP as the high-index medium for our simulations, our findings can be readily transposed to other materials, which might be more suitable for various applications.  Three such examples are diamond $\left(n = 2.4\right)$, SiC $\left(n = 2.5\right)$, and TiO$_2$ $\left(n = 2.5\right)$.  The lower refractive index contrast between the waveguiding region and the surrounding material lessens the confinement of the light in these cases and causes higher bending losses.  To compensate for these, the ring radius can be increased.  For example, for a diamond slot waveguide ring with $w = 180$~nm, $h = 230$~nm and $r = 3.1$~$\mu$m, we find that $Q = 30000$.  Likewise, for a SiC resonator with $w = 170$~nm, $h = 220$~nm and $r = 2.5$~$\mu$m, $Q = 29000$.  In other words, a change in the materials of the slot waveguide can be easily accounted for with a slight tuning of the geometry, allowing us to recover the functionality of our platform.

\subsection{Tunability}\label{sec:NanophotonicTuning}
The optical properties of slot waveguide rings can be changed by tuning the ring geometry (i.e. varying the size to alter $Q_{\rm rad}$) or by introducing losses, either due to enhanced scattering (e.g. due to surface roughness) or absorption.  While scattering is usually a static intrinsic feature, absorption losses can be introduced in either a passive or active manner, e.g., by doping the semiconductor layer during growth or actively by electrical or all-optical generation of free carriers.  Active control, in particular, allows for the reshaping of the emitted photon's wavefunction, if it occurs on the time-scale of the emitter lifetime~\cite{Active_control}.

\begin{figure*}[!htp]
  \centering
  \includegraphics[width=16.5cm]{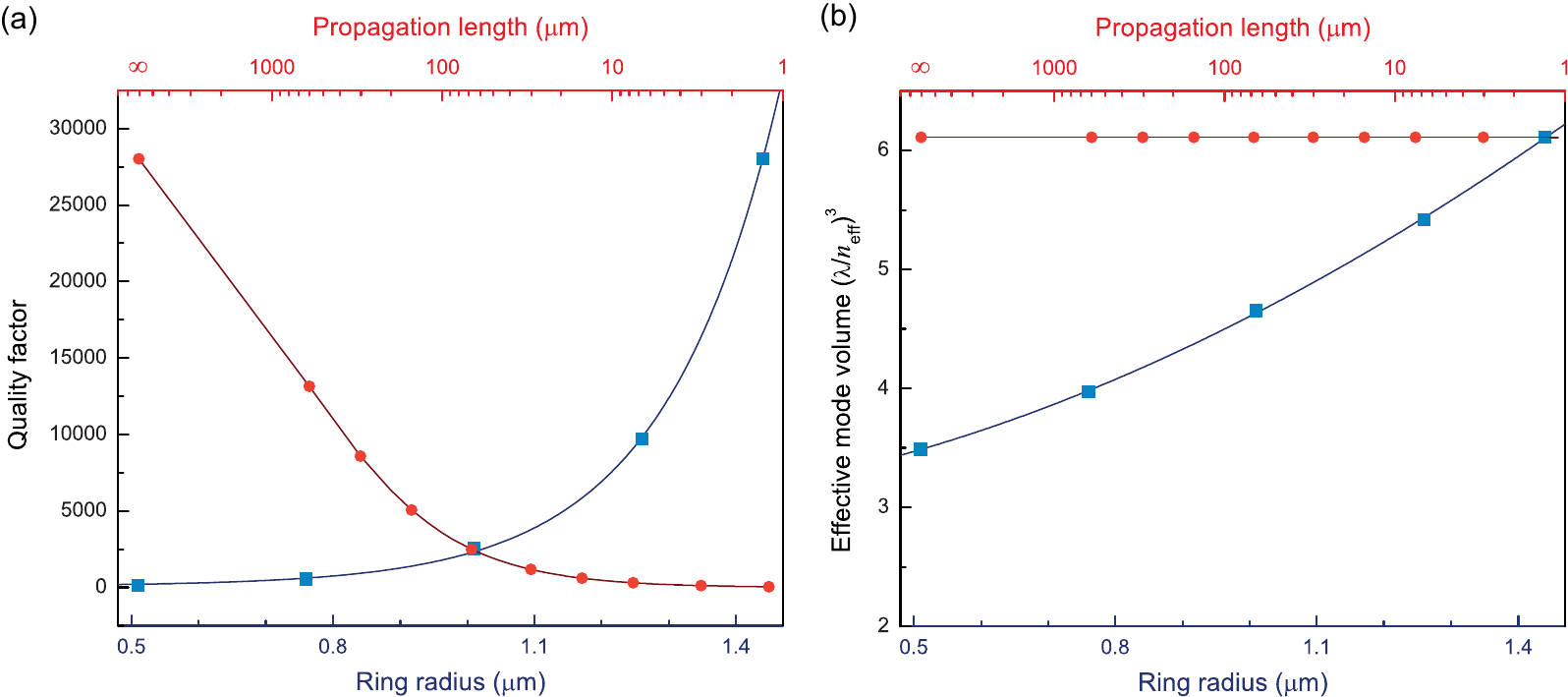}\\
  \vspace{-10 pt}\caption{Tuning slot waveguide ring properties. (a) Resonance quality factor as a function of ring radius (when $\kappa_{\rm GaP} = 0$,bottom axis) or the propagation length of light in GaP (when $r = 1.44$~$\mu$m, top axis).  The quality factor decreases monotonically as the ring shrinks, or as absorption in the GaP increases.  (b) The corresponding effective mode volume, $V_{\rm eff}$, for these rings, when the emitter with a radially-oriented dipole is located at the position marked in Fig.~\ref{fig:System}b.  Here, $V_{\rm eff}$ decreases as the ring shrinks, yet stays constant with increasing absorption. In both plots, the curves are guides for the eye.}\label{fig:QandVeff}
\end{figure*}

Here, we calculate the eigenmodes of rings (as was done in Fig.~\ref{fig:System}c), while varying either the ring radius or the imaginary part of the refractive index of GaP, $\kappa_{\rm GaP}$ to introduce absorption losses.  From these mode distributions and their corresponding complex eigenfrequencies, we extract both $Q$ and $V_{\rm eff}$~\cite{Mode_Volume_Formalism} as a function of the ring radius and the propagation length. The outcome is presented in Fig.~\ref{fig:QandVeff}.

As expected, both shrinking the ring and introducing absorption losses leads to a monotonic decrease in $Q$ (Fig.~\ref{fig:QandVeff}a), here by a factor of almost 300. These calculations show that even resonators with high losses (corresponding to propagation lengths in the 10's of micrometers) or with radii down to 1\,$\mu$m still have $Q$'s in the 1000's.

Interestingly, while $Q$ has a similar dependence on both the radius and the loss, changing these parameters has very different effects on $V_{\rm eff}$. Decreasing the ring size results in a smaller mode volume (bottom axis,  Fig.~\ref{fig:QandVeff}b) although in all cases we observe a small $V_{\rm eff}$ of only a few $\left(\lambda_{\rm eff}\right)^3$. This decrease, however, is not proportional to the decrease of the geometric ring volume, $V_{g} = \pi h \left(r_o^2 - r_i^2\right)$ where $r_{o,i}$ are the outer and inner radii of the ring.  For example, a decrease of $V_{g}$ by a factor of 2.8 results in a corresponding decrease of $V_{\rm eff}$ by only 1.75 (in all cases, $n_{\rm eff}$ ranges from 2.0 to 2.1).  This difference occurs because as the ring shrinks, the bending losses increase, and the mode is pushed out of the slot and into the outer part of the ring.  The situation is very different when $r$ is held constant and losses are ramped up.  In this case, the mode distribution is basically unaltered and $V_{\rm eff}$ remains constant (Fig.~\ref{fig:QandVeff}b, top axis).

\section{Implementations}
In this section we consider two different approaches to quantum optics in a slot waveguide ring.  In the first, quantum emitters are embedded at the electromagnetic mode maximum in the low-$n$ dielectric in the slot (see Fig.~\ref{fig:System}c). This approach is compatible with emitters such as single molecules and colloidal quantum dots, and it highlights the strong field confinement inherent to slot waveguides.  Secondly, we consider emitters embedded in the high-$n$ bars such as epitaxially grown quantum dots or defect centers in diamond. Here, we make use of the structured light fields of the slot waveguide ring to direct emission.

\subsection{Tunability of emission from low-$n$ quantum emitters}\label{sec:Imp_LowN}
\begin{figure*}[!htp]
  \centering
  \includegraphics[width=16.5cm]{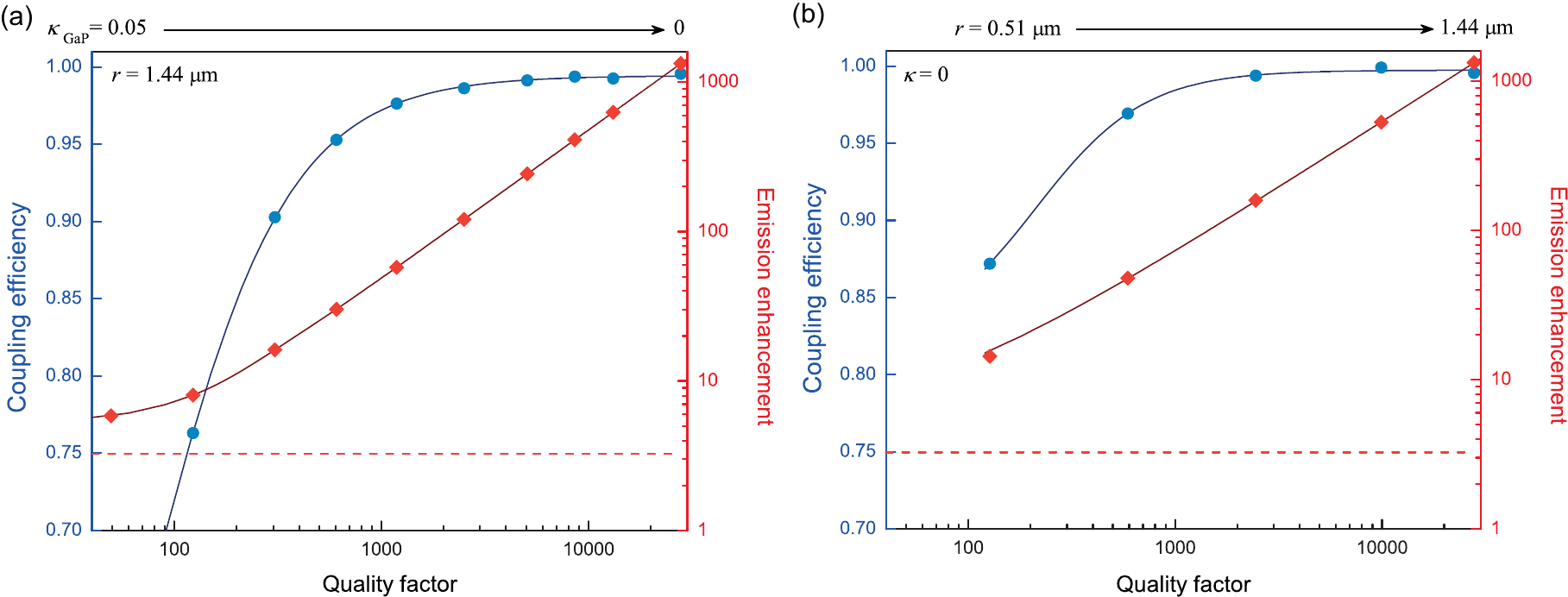}\\
  \vspace{-20 pt}\caption{Dipole emission into slot waveguide rings. Coupling efficiency $\beta$ (left axis) and emission rate enhancement $\chi$ (right axis) as a function of the resonance quality factor $Q$, which is changed by (a), introducing absorption into the GaP (same propagation lengths as in Fig.~\ref{fig:QandVeff}) and (b), changing the radius of the ring. In both plots, the performance of a lossless, straight slot waveguide is indicated by the dashed lines and the symbols are results of calculations while the solid curves are guides to the eye.}\label{fig:BetaEta}
\end{figure*}

In Sec.~\ref{sec:EmissionSWR} we saw that slot waveguide rings act as remarkably efficient interfaces between emitters and photons, when the emitters are placed in the field maximum inside the low-$n$ slots.  Here, we explore the tunability of the emission when the ring properties are varied, as was done in Sec.~\ref{sec:NanophotonicTuning}.  We begin by considering the case where absorption losses are introduced. In this case, $Q$ is changed while the optical eigenmode remains unaltered. The resultant emission properties are displayed in Fig.~\ref{fig:BetaEta}(a). Clearly, both $\beta$ and $\chi$ decrease as the losses increase (see Fig.~\ref{fig:QandVeff}a).  The ring, however, maintains its efficient performance for $\kappa_{\rm GaP}$ up to 0.004, resulting in $Q = 600$ and corresponding to a propagation length of 15~$\mu$m at $\lambda = 760$~nm. In this range, $\beta > 0.95$ while $\chi$ varies between 30 and 1,300.

Changing the ring radius instead of introducing losses, affects $\chi$ and $\beta$ in a similar way [Fig.~\ref{fig:BetaEta}(b)]: in both cases, these metrics decrease with decreasing $Q$. However, a close inspection of Fig.~\ref{fig:BetaEta} reveals a difference. The decrease in $\chi$ and $\beta$ is more rapid if $Q$ is lowered by absorption than if the lowering is caused by the increased radiation losses of the smaller rings because shrinking the ring results in a smaller mode volume, which counteracts the decrease in $Q$. In contrast, when absorption losses are introduced $Q$ decreases while the mode volume remains constant.  Since the emission properties depend on the ratio of $Q / V_{\rm eff}$~\cite{Nanophotonics}, we expect a smaller (lossless) ring to outperform a larger, lossy ring with a similar $Q$ as is most readily noticeable when $Q \lesssim 200$.

\subsection{Chiral emission with high-$n$ quantum emitters}\label{sec:Imp_highN}
Quantum emitters such as epitaxially grown quantum dots or vacancy centers in diamond, which are embedded in high-$n$ dielectrics, can also be interfaced with a slot waveguide ring.  As we briefly touched on in Sec.~\ref{sec:EmissionSWR}, placing the emitter away from the optical mode maximum necessarily results in a decrease to the emission rate enhancement.  For example, an emitter placed in one of the high-$n$ channels of a 1.44~$\mu$m radius ring, as shown by the blue square symbol in Fig.~\ref{fig:System}c, experiences $\chi = 56$ as compared to $\chi = 1,330$ at the mode maximum.  The emitter does, however, maintain $\beta = 0.99$ even when placed in one of the high-$n$ bars. We now show that at this position, the vectorial nature of light can be exploited to control the direction in which emission occurs. Such unidirectional coupling of emission to photonic pathways can be used as a basis for quantum architecture, and hence has been the focus of several recent studies~\cite{TNF_Directional, Directional_PhCW_dipoles, Directional_PhCW_QDsI, Directional_PhCW_QDsII}.

Unidirectional emission has been investigated for transitions between different spin-states of emitters, whose charge redistribution is described by circular dipoles (e.g. $\mathbf{d}_{e} = \sqrt{d/2}\left(\hat{r} \pm i \hat{\varphi}\right)$ in the ring coordinates).  For such directional emission to occur, the optical eigenmodes must contain regions of circular polarization, where the handedness of the light field depends on the direction of propagation.  Placing a circular dipole in such a region ensures that it would only radiate in one direction, depending on its handedness.  In Fig.~\ref{fig:Directionality}(a) we show the ellipticity (see Appendix~\ref{sec:Ellipticity}) of the light field in the $r = 1.44$~$\mu$m ring [whose mode is shown in Fig.~\ref{fig:System}(c)]. This quantity is a measure for how circular the light field is, peaking at $\pm 1$ where the light is right or left handed circularly polarized, while for linear light fields it is 0.

\begin{figure}[!htp]
  \centering
  \includegraphics[width=8cm]{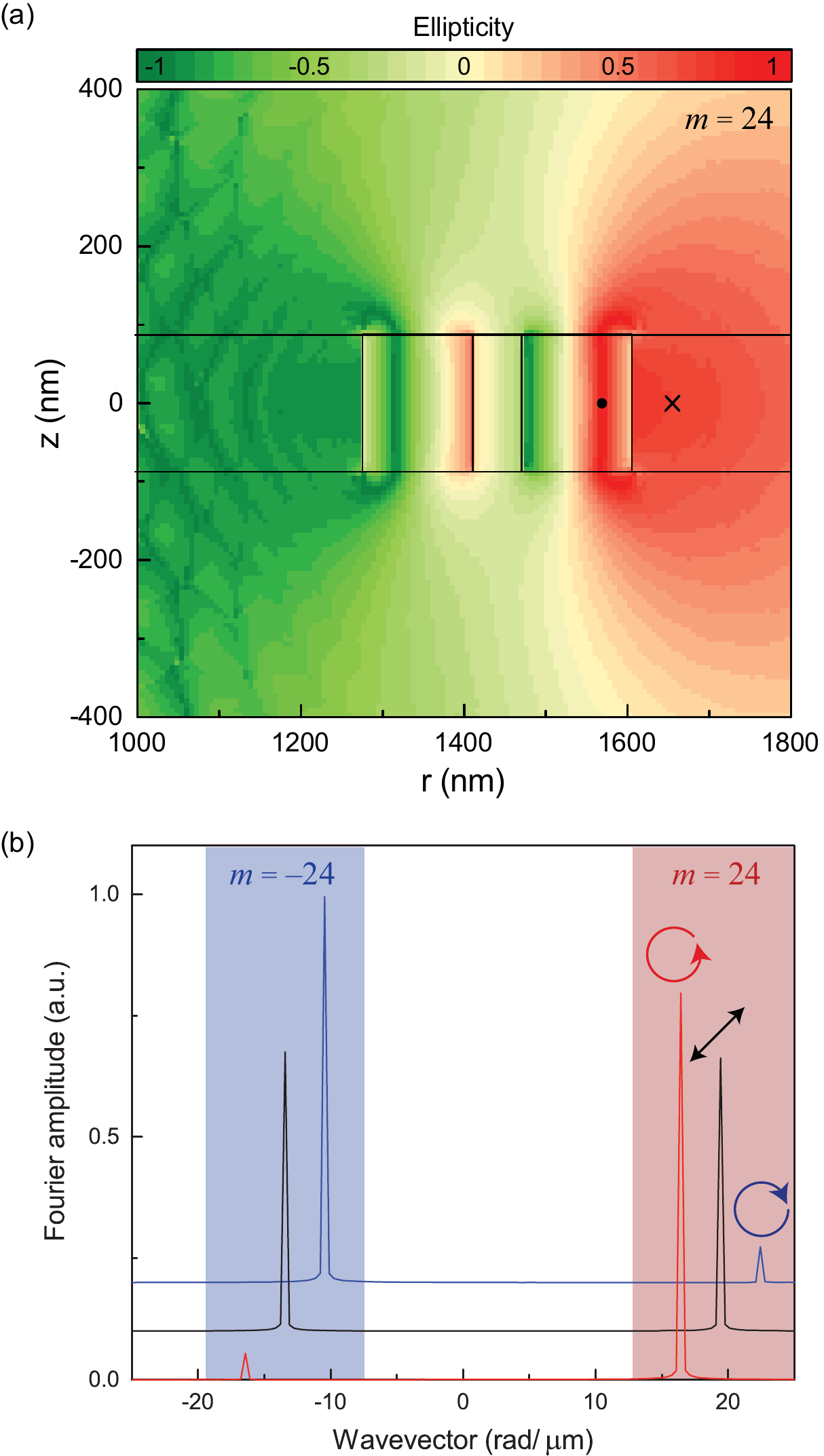}\\
  \caption{Directional coupling of emission into a slot waveguide ring.  (a) The ellipticity of the light field in the slot waveguide ring, where values of $\pm 1$ correspond to right and left handed circular polarization, and 0 corresponds to linearly polarized light. Symbols indicate possible locations of dipole emitters, as discussed in the text.  (b)  The Fourier amplitudes for a line trace of $\mbox{Im}E_{r}$ at the center of the slot, radiated by differently oriented dipoles that are placed at the position shown by the solid circle in (a).  In all cases we observe two peaks at $\pm 16.4$~rad$/\mu$m, corresponding to the $\pm 24$ azimuthal-order modes.  The relative size of each peak corresponds to the emission magnitude into the corresponding mode. The respective dipole orientation is shown above the curves, which have been shifted in both dimensions for clarity.}\label{fig:Directionality}
\end{figure}

From Fig.~\ref{fig:Directionality}(a) it is clear that the areas where the light is most circular can be found outside the slot.  In fact, since we want both a near-unity ellipticity, as well as a large field amplitude, a favorable position for directional emission is inside the high-index bars (solid circle in Fig.~\ref{fig:Directionality}a); at this position the ellipticity peaks at 0.87.

We repeat our calculations with an emitter placed at the position marked by the dot in Fig.~\ref{fig:Directionality}(a).  We vary the transition dipole to consider linear as well as right and left handed circularly polarized dipoles.  We Fourier transform the line-trace of $E_r$ along the center of the slot to obtain the wavenumber spectrum, whose amplitude is shown in Fig.~\ref{fig:Directionality}(b).  In this transform, we observe two sharp peaks, centered about $\pm 16.4$~rad$/\mu$m, corresponding to light propagating in the $m = \pm 24$ modes, respectively.  By comparing the area under these peaks, we determine the directionality of the emission.  In the case of the linear dipole (black curve), the two peaks are almost identical and there is no directionality to within a $2\%$ calculation error.  In contrast, for the two circular dipoles (blue and red curves), one peak in each curve dominates, depending on the handedness of the dipole.  We obtain a directionality of $0.87 \pm 0.02$, as expected from the ellipticity of the mode.  For completeness, we also calculate the situation of a circular dipole placed in the low index material outside the slot waveguide (cross in Fig.~\ref{fig:Directionality}a), finding a directionality of $0.75 \pm 0.02$.

A slot-waveguide ring, therefore, ideally lends itself as an element in a chiral quantum network~\cite{QuantumGates}.  An emitter such as a quantum dot would experience $\beta = 0.99$, which can be decomposed to $\beta_+ = 0.86$ and $\beta_- = 0.13$ for the two counter-propagating modes.  We expect that slight adjustments to the waveguide geometry would increase the ellipticity close to unity, and hence allow for perfect directional emission.  Finally, we note that the $\chi = 56$ calculated for such an emitter would, for example, sufficiently broaden the emission spectrum and overcome residual line broadenings often encountered in the solid state.

\section{Conclusions}
In this work we introduced the use of slot waveguide rings for quantum optics.  We combined numerical models with analytic quantum theory to study the emission properties of single emitters coupled to the rings. We demonstrated that using rings that can realistically be fabricated, with a geometric footprint as small as 6.5 $\mu$m$^2$, it is possible to strongly couple a solid-state quantum emitter such as an organic dye molecule to the photonic modes.  In particular, Rabi oscillations can be clearly visible in the decay dynamics of such moderate $Q$ rings.

We also showed that it is possible to tune emission properties by changing the optical mode of the slot waveguide rings.  Two different tuning mechanisms were identified: a change of the ring size or the introduction of absorption to the system.  While the former is a passive effect, active control of the later at speeds up to ultrafast time scales offers a fascinating gateway to all-optical control of complex nanophotonic interactions.  In either case, tuning the system led to $\beta$ values ranging from 0.75 to near unity and $\chi$ spanning three orders of magnitude up to about 1300.

In closing we discussed the unique potential of slot waveguide rings for unidirectional emission.  As examples, positions both inside the high index bars and in the low index medium outside the slot waveguide ring, were identified.  Interestingly, strongly directional emission can occur at positions where $\beta > 0.99$ and $\chi = 56$. In summary, the combination of efficient emission enhancement with a high degree of tunability suggest a highly attractive platform for both investigations of fundamental quantum phenomena, and for future quantum optics technology.

\appendix
\section{Numerical calculations of $\bm{\chi}$ and $\bm{\beta}$}\label{sec:etaBeta}
The power dissipated by the radiation of a dipole is proportional to the out of phase-response of the electric field at its position~\cite{Nanophotonics}.  Specifically, the power dissipated by a dipole (with dipole moment $\mathbf{d}_{e}$) located at point $\mathbf{r}_{e}$ and radiating at a frequency $\omega_e$ is
\begin{equation}\label{eq:Dis}
    P=\frac{\omega_{e}}{2}\mbox{Im}\left[\mathbf{d}_{e}\cdot\mathbf{E}\left(\mathbf{r}_{e}\right)\right],
\end{equation}
where $\mathbf{E}\left(\mathbf{r}\right)$ is the field radiated by the dipole.  In our scenario, we assume radially oriented transition dipoles and extract the out-of-phase component $\mbox{Im}E_{r}$ from simulations.  We can then calculate $\chi$ and $\beta$, which both characterize emission properties in the presence of the nanophotonic structure, and are used in our analytic model to describe the decay dynamics of the emitter.

\begin{figure*}[!htp]
  \centering
  \includegraphics[width=16.5cm]{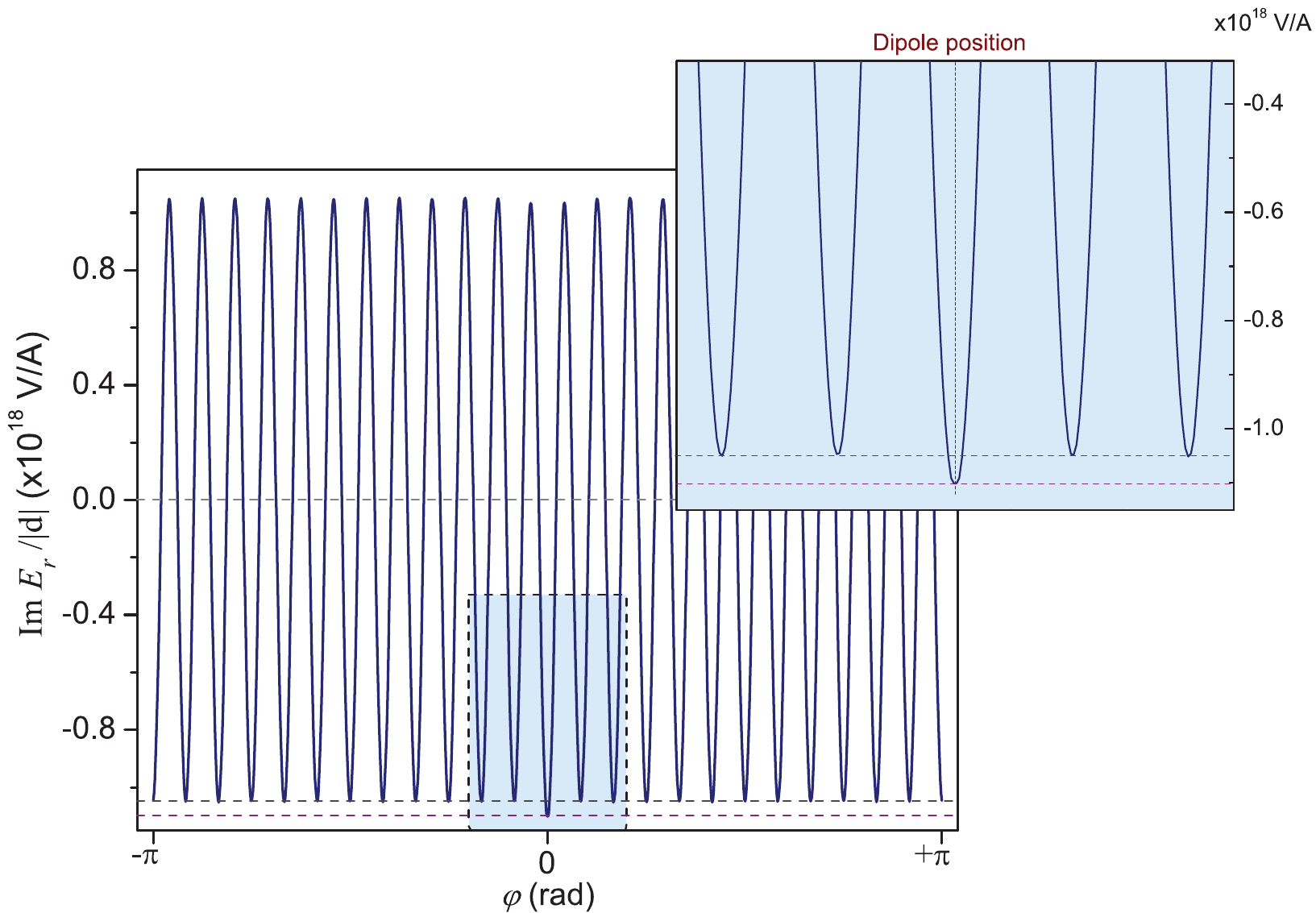}\\
  \vspace{-10 pt}\caption{Out-of-phase component of the radial field at a constant distance, corresponding to the position of the emitter, along the ring for a $r=1.44$~$\mu$m, $\kappa = 0.004$ ring.  The dipole source is radially oriented. The inset shows a zoom-in on the position of the dipole source, at $\varphi = 0$.	In both figures we mark the field at the position of the dipole (dashed, purple line) and in the steady state along the ring (dashed, dark gray line).}\label{fig:ImEr}
\end{figure*}

The emission enhancement  $\chi$  is the ratio of the power dissipated by the radiation of a dipole in a nanophotonic structure (subscript `nano') to that of the same dipole, but in bulk media (subscript `hom').  That is, for a linear transition dipole, Eq.~\ref{eq:Dis} allows us to write,
\begin{equation}\label{eq:eta}
   \chi=\frac{\mbox{Im}E_{r, {\rm \, nano}}\left(\mathbf{r}_{e}\right)}{\mbox{Im}E_{r, \, {\rm hom}}\left(\mathbf{r}_{e}\right)}.
\end{equation}
We find $\mbox{Im}E_{r, {\rm \, nano}}\left(\mathbf{r}_{e}\right)$, by looking at a line trace from the results of the 3D simulations (e.g. Fig.~\ref{fig:Emission}), as shown in Fig.~\ref{fig:ImEr}.
For this ring, $\mbox{Im}E_{r, \, {\rm nano}}\left(\mathbf{r}_{0}\right) / \left|d\right|= -1.11 \times 10^{18}$~V/A.  Similarly, for this dipole in bulk naphthalene, we find (using simulations in a half-spherical space, as in the case of the slot waveguide ring) that  $\mbox{Im}E_{r, \, {\rm hom}}\left(\mathbf{r}_{0}\right) / \left|d\right| = -4.55 \times 10^{15}$~V/A.  Hence, for this particular ring, $\chi = 243$, as reported in the main text.  For completeness, we note that in the weak coupling limit, we can express $\chi$ in terms of the decay rates of the emitter, writing $\chi = \gamma_{\rm{nano}} / \gamma_{\rm{hom}}$; in the single mode limit this is equivalent to the Purcell Factor.~\cite{Mode_Volume_Formalism}

To find how well the molecule emits into the desired mode of the ring, we compare $\mbox{Im}E_{r}$ at the position of the dipole to that in the far-field of the emitter, but at the same radial distance.  This comparison can be seen in Fig.~\ref{fig:ImEr}, corresponding to the values shown by the dashed purple and dark-gray lines.  For this ring ($\kappa = 0.004$) this difference is clearly visible, and it corresponds to $\beta = 0.949$.

When large losses are introduced into the system, either due to absorption in the GaP bars or due to increased radiation in small rings, we need to take an extra step to correctly calculate $\beta$.  In the presence of large losses, the radiated field decays away from its source, meaning that there is no constant line-trace to the far-field $\mbox{Im}E_{r}\left(\mathbf{r}_{0}\right)$, as was the case in Fig.~\ref{fig:ImEr}.  In this situation, we fit a decaying exponential envelop function to the field trace using only the far-field field amplitude.  The extrapolated value of this envelope function, at the position of the dipole, is then used to calculate $\beta$.

\section{Analytic expression for the $C_e\left(t\right)$}\label{sec:Theory}
To understand the dynamics of our system we calculate the time-dependent probability to find an emitter inside our slot waveguide ring in its excited state.  We develop this expression using the framework of macroscopic quantum electrodynamics (QED)~\cite{Theory_dA2Gamma} and follow the approach of Khanbekyan \emph{et al.}~\cite{Emitter_Cavity_Theory}.

The Hamiltonian of our system, which consists of a single, two-level emitter and, at most, a single excitation of the slot waveguide ring is
\begin{equation}\label{eq:H}
    \hat{H}=\int d^{3}\mathbf{r}\int_{0}^{\infty}d\omega\hbar\hat{\mathbf{f}}^{\dagger}\left(\mathbf{r},\omega\right)\cdot\hat{\mathbf{f}}\left(\mathbf{r},\omega\right)+\hbar\omega_{e}\hat{S}_{+}\cdot\hat{S}_{-}-\hat{\mathbf{d}}_{e}\cdot\hat{\mathbf{E}}\left(\mathbf{r}_{e}\right).
\end{equation}
Here, the first term is the free Hamiltonian of the electromagnetic field inside the structure, with $\hat{\mathbf{f}}^{\dagger}$ and $\hat{\mathbf{f}}$ being the bosonic creation and annihilation operators.  Likewise, the second term is the free Hamiltonian of the emitter, where $\hat{S}_{+}=\left|e\right\rangle \left\langle g\right|$ and $\hat{S}_{-}=\left|g\right\rangle \left\langle e\right|$ are the pseudo spin-flip operators that correspond to transitions between the excited and ground states.  The third term is the interaction Hamiltonian, which represents the emitter-field coupling in the dipole approximation, where the emitter is described by its dipole operator
\begin{equation}\label{eq:d}
    \hat{\mathbf{d}}_{e} = \mathbf{d}_{e}\cdot\left(\hat{S}_{+}+\hat{S}_{-}\right),
\end{equation}
and the field is given by
\begin{equation}\label{eq:E}
    \hat{\mathbf{E}}\left(\mathbf{r}\right)=i\sqrt{\frac{\hbar}{\varepsilon_{0}\pi}}\int_{0}^{\infty}d\omega\frac{\omega^{2}}{c^{2}}\int d^{3}\mathbf{r}'\sqrt{\varepsilon''\left(\mathbf{r}',\omega\right)}\mathrm{G\left(\mathbf{r},\mathbf{r}',\omega\right)\cdot\hat{\mathbf{f}}\left(\mathbf{r}',\omega\right)}+\mbox{h.c.},
\end{equation}
where $\mathrm{G}\left(\mathbf{r},\mathbf{r}',\omega\right)$ is the classical electromagnetic Green's function of the nanostructure, and $\varepsilon''\left(\mathbf{r}',\omega\right)$ is the position and frequency dependent, imaginary component of the relative permittivity of our structure.

The wavefunction in the single-excitation limit is
\begin{equation}\label{eq:Wavefunction}
    \left|\psi\left(t\right)\right\rangle =C_{e}\left(t\right)e^{-i\omega_{e}t}\left|e, 0\right\rangle +\int d^{3}\mathbf{r}\int_{0}^{\infty}d\omega C_{g}\left(\mathbf{r},\omega,t\right)e^{-i\omega t}\left|g, 1\left(\mathbf{r},\omega\right)\right\rangle.
\end{equation}
Here, the basis states of the system are $\left|e, 0\right\rangle$ where the emitter is excited and there is no photon, and $\left|g, 1\left(\mathbf{r},\omega\right)\right\rangle$ where the emitter is in its ground state and there is a single photon.  In this equation, $\omega_{e}$ is the transition frequency of our emitter, and the complex coefficients $C_{e}\left(t\right)$ and $C_{g}\left(\mathbf{r},\omega,t\right)$ can be used to find the time-dependent probabilities for the system to be in each state.

We use this wavefunction when solving Schr{\"o}dinger's equation,
\begin{equation}\label{eq:SE}
    i\hbar\partial_{t}\left|\psi\left(t\right)\right\rangle =\hat{H}\left|\psi\left(t\right)\right\rangle,
\end{equation}
in the rotating wave approximation, where only the resonant terms in the interaction Hamiltonian survive (i.e. those that contain either $\hat{S}_{+}\hat{\mathbf{f}}$ or $\hat{S}_{-}\hat{\mathbf{f}}^{\dagger}$).  We compare the two sides of Eq.~\ref{eq:SE} to arrive at the following differential equation for the coefficients of Eq.~\ref{eq:Wavefunction},
\begin{eqnarray}
\label{eq:dCe}\dot{C}_{e}\left(t\right)	&=&	-\frac{\mathbf{d}_{e}}{\sqrt{\varepsilon_{0}\hbar\pi}}\int_{0}^{\infty}d\omega\frac{\omega^{2}}{c^{2}}\int d^{3}\mathbf{r}\sqrt{\varepsilon''\left(\mathbf{r},\omega\right)}\mathrm{G\left(\mathbf{r}_{e},\mathbf{r},\omega\right)}C_{g}\left(\mathbf{r},\omega		 ,t\right)e^{-i\left(\omega-\omega_{e}\right)t},\\
\label{eq:dCg}\dot{C}_{g}\left(z,\omega,t\right)&=&\frac{\mathbf{d}_{e}^{*}}{\sqrt{\varepsilon_{0}\hbar\pi}}\frac{\omega^{2}}{c^{2}}\sqrt{\varepsilon''\left(\mathbf{r},\omega\right)}\mathrm{G^{*}\left(\mathbf{r}_{e},\mathbf{r},\omega\right)}C_{e}\left(t\right)e^{i\left(\omega-\omega_{e}\right)t}.
\end{eqnarray}

To proceed, we recall that the molecule is, initially, in its excited state, meaning that $C_{g}\left(\mathbf{r},\omega,0\right)=0$.  Inserting this expression into Eq.~\ref{eq:dCg}, and using the fundamental theorem of calculus allows us to write
\begin{equation}\label{eq:Cg}
    C_{g}\left(\mathbf{r},\omega,t\right)=\frac{\mathbf{d}_{e}^{*}}{\sqrt{\varepsilon_{0}\hbar\pi}}\frac{\omega^{2}}{c^{2}}\sqrt{\varepsilon''\left(\mathbf{r},\omega\right)}\mathrm{G}^{*}\left(\mathbf{r}_{e},\mathbf{r},\omega\right)\int_{0}^{t}dt'C_{e}\left(t'\right)e^{i\left(\omega-\omega_{e}\right)t'}.
\end{equation}
Next, we make use of the fact that the modes of our slot waveguide ring are semi-discrete, meaning that for every mode $k$ the FWHM $\Gamma_{k}$ is much shorter than the free spectral range.  In this case, we may safely model the Green's function by a sum of Lorentzian resonances,
\begin{equation}\label{eq:Gdiscrete}
    \frac{\omega^{2}}{c^{2}}G\left(z_{e},z,\omega\right)\cong\sum_{k}\frac{\Omega_{k}^{2}}{c^{2}}\tilde{G}\left(z_{e},z,\Omega_{k}\right)\frac{-i \Gamma_{k}/2}{\Omega_{k}-\omega-i\Gamma_{k}/2},
\end{equation}
where $\tilde{G}\left(z_{A},z,\Omega_{k}\right)$ is the Green's function amplitude at the central frequency, $\Omega_{k}$, of mode $k$.

We can then insert Eq.~\ref{eq:Cg} into Eq.~\ref{eq:dCe}, and use Eq.~\ref{eq:Gdiscrete} to perform the frequency integral for a single mode, which allows us to write
\begin{equation}\label{eq:CeK}
    \dot{C}_{e}\left(t\right)=\int_{0}^{t}dt'K\left(t-t'\right)C_{e}\left(t'\right),
\end{equation}
where the time-dependent kernel is
\begin{equation}\label{eq:K}
    K\left(t\right) = K_{0}e^{-i\left(\omega_{\rm cav}-\omega_{e}\right)t}e^{-\frac{\gamma_{\rm cav}}{2}t},
\end{equation}
and,
\begin{equation}\label{eq:K0_G}
    K_{0}=-\frac{\left|\mathbf{d}_{e}\right|^{2}}{\varepsilon_{0}\hbar}\frac{\omega_{\rm cav}^{2}}{c^{2}}\frac{\gamma_{\rm cav}}{2}\mbox{Im}\tilde{G}\left(\mathbf{r}_{e},\mathbf{r}_{e},\omega_{\rm cav}\right).
\end{equation}
Here, we have used the following Green's function identity~\cite{Theory_dA2Gamma},
\begin{equation}\label{eq:Gident}
    \mbox{Im}\mathrm{G}\left(z_{1},z_{2},\omega\right)=\frac{\omega^{2}}{c^{2}}\int dz\varepsilon''\left(z,\omega\right)\mathrm{G}\left(z_{1},z,\omega\right)\mathrm{G}^{*}\left(z_{2},z,\omega\right),
\end{equation}
and assumed that the molecule only interacts with the $k=0$ mode of the structure.

We now take the time derivative of Eq.~\ref{eq:CeK}, and then make use of Eq.~\ref{eq:K}, to write
\begin{equation}\label{eq:ddCe}
    \ddot{C}_{e}\left(t\right)+\Gamma\dot{C}_{e}\left(t\right)-K_{0}C_{e}\left(t\right)=0,
\end{equation}
where $\Gamma=i\left(\omega_{\rm cav}-\omega_{e}\right)+\gamma_{\rm cav}/2$, as defined in the main text.  To solve this second-order differential equation we impose initial conditions of the system.  That is, since the molecule is initially excited, $C_{e}\left(0\right)=1$ and $\dot{C}_{e}\left(0\right)=0$.  The latter condition follows from Eq.~\ref{eq:dCg}, and that if $C_{e}\left(0\right)=1$ then $C_{g}\left(\mathbf{r},0\right)=0$.  The solution provides Eq.~\ref{eq:Ce} of the main text.

Lastly, we show how we are able to rewrite Eq.~\ref{eq:K0_G} in the form of $K_0$ from the main text.  First, we note that in Eq.~\ref{eq:K0_G} we do not know the value of the dipole moment of the emitter,  $\mathbf{d}_{e}$.  We do, however, know the experimentally measured linewidth of the emitter in a bulk environment, $\gamma_{\rm {hom}}$.  This linewidth can be related to the Green's function of the bulk media by~\cite{Theory_dA2Gamma},
\begin{equation}\label{eq:GammaHom}
    \gamma_{\rm {hom}}=\frac{2\left|\mathbf{d}_{e}\right|^{2}}{\varepsilon_{0}\hbar}\frac{\omega_{e}^{2}}{c^{2}}\mbox{Im}\tilde{G}_{\rm {hom}}\left(\mathbf{r}_{e},\mathbf{r}_{e},\omega_{e}\right),
\end{equation}
where $\tilde{G}_{\rm {hom}}\left(\mathbf{r}_{e},\mathbf{r}_{e},\omega_{e}\right)$ can be either calculated analytically~\cite{Nanophotonics} or extracted from simulations.  Thus, we can express the dipole moment via the linewidth and write
\begin{equation}\label{eq:K0_Gratio}
    K_{0}=-\frac{\omega_{\rm cav}^{2}}{\omega_{e}^{2}}\frac{\gamma_{\rm cav}\gamma_{\rm {hom}}}{4}\frac{\mbox{Im}\tilde{G}\left(\mathbf{r}_{e},\mathbf{r}_{e},\omega_{\rm cav}\right)}{\mbox{Im}\tilde{G}_{\rm {hom}}\left(\mathbf{r}_{e},\mathbf{r}_{e},\omega_{e}\right)}.
\end{equation}
Since the electric field radiated by a dipole defines the Green's function,
\begin{equation}\label{eq:E_dipole}
    \mathbf{E}\left(\mathbf{r},\omega\right)=\frac{\omega^{2}}{c^{2}\varepsilon_{0}}\mathrm{G\left(\mathbf{r},\mathbf{r}_{e},\omega\right)}\cdot\mathbf{\hat{d}}_{e},
\end{equation}
we can rewrite Eq.~\ref{eq:eta} as
\begin{equation}\label{eq:etaG}
    \chi = \frac{\mbox{Im}\tilde{G}\left(\mathbf{r}_{e},\mathbf{r}_{e},\omega_{\rm cav}\right)}{\mbox{Im}\tilde{G}_{\rm {hom}}\left(\mathbf{r}_{e},\mathbf{r}_{e},\omega_{e}\right)}.
\end{equation}
Finally, using Eq.~\ref{eq:etaG} in Eq.~\ref{eq:K0_Gratio} allows us to arrive at the expression for $K_0$ that is given in the main text.

\section{Ellipticity}\label{sec:Ellipticity}
If we consider the polarization state of a near field then at every point in space the electric field vector tip will trace out an ellipse as time passes and it rotates.  The ellipticity of the light field is the ratio of the short to the long axis of this polarization ellipse and, in our case where only two components of the electric field are non-zero (i.e. $E_{r, \varphi} \neq 0$ while $E_z \simeq 0$, over most space as is shown in Fig.~\ref{fig:System}(b)), can be written as
\begin{equation}\label{eq:Ellipticity}
    \epsilon\left(\mathbf{r}\right) = \tan\left\{ \sin^{-1}\left(\sin\left[2\psi\left(\mathbf{r}\right)\right]\sin\left[\delta\left(\mathbf{r}\right)\right]\right)/2\right\}.
\end{equation}
Here $\psi\left(\mathbf{r}\right) = \tan^{-1}\left[\left|E_{\varphi}\left(\mathbf{r}\right)\right| / \left|E_r\left(\mathbf{r}\right)\right|\right]$ and $\delta\left(\mathbf{r}\right) = \delta_{\varphi}\left(\mathbf{r}\right) - \delta_r\left(\mathbf{r}\right)$ are the ratio and difference of the field amplitudes and phases, respectively.  The ellipticity ranges from $-1$ to $+1$, where it is left- and right-circularly polarized, respectively.  When $\epsilon = 0$, the light is linearly polarized.

\bibliography{Rotenberg}

\end{document}